  \def\versionno{ duality -- version 1.0 }

\catcode`\@=11

\catcode`\@=11

\expandafter\ifx\csname  
draftcontrol\endcsname\relax\global\def\draftcontrol{0}
\fi

{\count255=\time\divide\count255 by 60
\xdef\hourmin{\number\count255}
\multiply\count255 by-60\advance\count255 by\time
\xdef\hourmin{\hourmin:\ifnum\count255<10 0\fi\the\count255}}
\def\draftdate{\number\month/\number\day/\number\year\ \ \ \hourmin }

\newcommand\makepapertitle{\par
   \begingroup
     \renewcommand\thefootnote{\@fnsymbol\c@footnote}%
     \def\@makefnmark{\rlap{\@textsuperscript{\normalfont\@thefnmark}}}%
     \long\def\@makefntext##1{\parindent 1em\noindent
             \hb@xt@1.8em{%
                 \hss\@textsuperscript{\normalfont\@thefnmark}}##1}%
      \newpage
      \global\@topnum\z@   
      \@makepapertitle
      \thispagestyle{empty}\@thanks
   \endgroup
   \setcounter{footnote}{0}%
   \global\let\thanks\relax
   \global\let\makepapertitle\relax
   \global\let\@makepapertitle\relax
   \global\let\@thanks\@empty
   \global\let\@author\@empty
   \global\let\@date\@empty
   \global\let\@title\@empty
   \global\let\title\relax
   \global\let\author\relax
   \global\let\date\relax
   \global\let\and\relax
   \def\version{\let\version\@version\@gobble}
}
\def\@makepapertitle{%
   \newpage
    \ifnum\draftcontrol=1 {}
    \version\versionno
    \vskip 3em%
    \else
    \hfill\hbox to 3cm {\parbox{4cm}{\@pubnum}\hss}%
    \vskip 3em%
    \fi
    \begin{center}%
    \let \footnote \thanks
      {\LARGE \@title \par}%
      \vskip 1.5em%
      {\normalsize
        \lineskip .5em%
        \begin{center} 
          \@author
        \end{center} 
\par}%
      \vskip 1em%
      {\@bstract}%
      \end{center}%
      \vskip .5em
      \@date%
    \par
}

\gdef\@pubnum{}
\def\pubnum#1{%
   \gdef\@pubnum{#1}}

\gdef\@bstract{}
\def\Abstract#1{%
   \gdef\@bstract{%
    \parbox{\textwidth-0pc}{%
    \centerline{\bf Abstract}\penalty1000
    \noindent
    \renewcommand\baselinestretch{1.0}
    {#1}}}
}

\def\ps@paper{\let\@mkboth\@gobbletwo%
      \ifnum\draftcontrol=1
         \def\@oddfoot{\hbox to \textwidth{\tiny \versionno  
\hfil\tiny\draftdate}%
         \hskip -\textwidth \hbox to \textwidth{\hfil\rm\thepage\hfil}}%
      \else\def\@oddfoot{\hbox to \textwidth{\hfil\rm\thepage\hfil}}
      \fi
      \let\@evenfoot\@oddfoot
}

\def\body{\clearpage
           \pagestyle{paper}
         }
\newenvironment{acknowledgments}{%
\vskip 3.25ex
\noindent {\bf Acknowledgments}
}

\def\@version#1{\ifnum\draftcontrol=1
\typeout{}\typeout{#1}\typeout{}
\vskip3mm\centerline{\hbox{\fbox{\normalsize{\tt DRAFT -- #1 -- }
                    {\draftdate}}}}\vskip3mm
\fi}
\let\version\@version
\long\def\eqlabel#1{\ifnum\draftcontrol=1
                     \tag@false  
without this
                     \tag*{(\theequation) \hbox to  
-0.2cm{\hspace{0cm}\small{#1}\hss}}
                     \refstepcounter{equation}
                     \edef\@currentlabel{\theequation}
                     \ltx@label{#1}          
instead of new definition
                     \else
                     \label{#1}
                     \fi
                     }
\let\st@bibitem\@bibitem
\let\st@lbibitem\@lbibitem
\ifnum\draftcontrol=1
   \def\@bibitem#1{%
     \st@bibitem{#1}\a@@label{#1}\ignorespaces}
   \def\@lbibitem[#1]#2{%
     \st@lbibitem[#1]{#2}\a@@label{#2}\ignorespaces}
   \def\a@@label#1{%
     \gdef\a@lab{\smash{\normalfont\small#1}}
     \ifvmode
       \if@inlabel
         \global\setbox\@labels\hbox{%
           \llap{\a@lab\let\a@lab\relax
                 \kern\@totalleftmargin\kern\marginparsep}%
           \box\@labels}%
       \fi
     \fi}
\fi

\documentclass[12pt,letterpaper]{article}

\usepackage{amsmath,amssymb,array,calc,rotating,epsfig,psfrag}
\usepackage{hyperref}

\ifnum\draftcontrol=1
\fi

\renewcommand\baselinestretch{1.25}
\renewcommand\section{\@startsection {section}{1}{\z@}%
                                    {-3.5ex \@plus -1ex \@minus -.2ex}%
                                    {2.3ex \@plus.2ex}%
                                    {\normalfont\large\bfseries}}
\renewcommand\subsection{\@startsection{subsection}{2}{\z@}%
                                      {-3.25ex\@plus -1ex \@minus -.2ex}%
                                      {1.5ex \@plus .2ex}%
                                      {\normalfont\normalsize\bfseries}}
\renewcommand\subsubsection{\@startsection{subsubsection}{3}{\z@}%
                                      {-3.25ex\@plus -1ex \@minus -.2ex}%
                                      {1.5ex \@plus .2ex}%
                                      {\normalfont\normalsize\it}}

\numberwithin{equation}{section}

\def\complex      {{\mathbb C}}

\def\projective   {{\mathbb P}}

\def\reals        {{\mathbb R}}
\def\zet          {{\mathbb Z}}

\def\revise#1       {\marginpar{\rule{2mm}{1cm} #1}}

\def\ZZ{\zet}
\def\RR{\reals}
\def\PP{\projective}
\def\RP{\RR\PP}
\def\N{\caln}

\def\R{{\rm R}}

\def\sqr#1#2{{\vcenter{\vbox{\hrule height.#2pt
  \hbox{\vrule width.#2pt height#1pt \kern#1pt
  \vrule width.#2pt}\hrule height.#2pt}}}}

\def\yboxit#1#2{\vbox{\hrule height #1 \hbox{\vrule width #1
\vbox{#2}\vrule width #1 }\hrule height #1 }}
\def\fillbox#1{\hbox to #1{\vbox to #1{\vfil}\hfil}}
\def\ybox{{\lower 1.3pt \yboxit{0.4pt}{\fillbox{8pt}}\hskip-0.2pt}}

\def\comments#1{}

\def\CC{{\cal C}}

\def\CN{{\cal N}}

\def\P{\BP}

\def\II{\relax{I\kern-.10em I}}

\def\IZ{\relax\ifmmode\mathchoice
{\hbox{\cmss Z\kern-.4em Z}}{\hbox{\cmss Z\kern-.4em Z}}
{\lower.9pt\hbox{\cmsss Z\kern-.4em Z}}
{\lower1.2pt\hbox{\cmsss Z\kern-.4em Z}}\else{\cmss Z\kern-.4em
Z}\fi}
\def\IB{\relax{\rm I\kern-.18em B}}
\def\IC{{\relax\hbox{$\inbar\kern-.3em{\rm C}$}}}
\def\ID{\relax{\rm I\kern-.18em D}}
\def\IE{\relax{\rm I\kern-.18em E}}
\def\IF{\relax{\rm I\kern-.18em F}}
\def\IG{\relax\hbox{$\inbar\kern-.3em{\rm G}$}}
\def\IGa{\relax\hbox{${\rm I}\kern-.18em\Gamma$}}
\def\IH{\relax{\rm I\kern-.18em H}}
\def\II{\relax{\rm I\kern-.18em I}}
\def\IK{\relax{\rm I\kern-.18em K}}
\def\IP{\relax{\rm I\kern-.18em P}}

\def\inbar{\,\vrule height1.5ex width.4pt depth0pt}

\font\cmss=cmss10 \font\cmsss=cmss10 at 7pt
\def\IR{\relax{\rm I\kern-.18em R}}

\def\BP{\IP}

\def\lp10{l_P^{10}}
\def\lp11{l_P^{11}}

\newcommand{\nc}{\newcommand}
\nc{\rnc}{\renewcommand}
\nc{\CY}{Calabi-Yau}
\nc{\CYM}{Calabi-Yau manifold}
\nc{\CYMs}{Calabi-Yau manifolds}
\nc{\DB}{D-Brane}
\nc{\DBs}{D-Branes}
\nc{\SUSY}{supersymmetry}
\nc{\Kah}{K\"ahler}
\nc{\cs}{complex structure}
\nc{\beq}{\begin{equation}}
\nc{\eeq}{\end{equation}}
\nc{\beqa}{\begin{eqnarray}}
\nc{\eeqa}{\end{eqnarray}}
\nc{\ntwo}{${\cal N}=2$}
\nc{\nOne}{${\cal N}=1$}
\nc{\hs}{\hspace{0.2in}}
\nc{\Z}{{\mathbb Z}}
\rnc{\P}{{\mathbb P}}
\rnc{\RP}{{\mathbb {RP}}}
\nc{\WP}{\mathbb{WP}}
\nc{\slag}{special Lagrangian}
\nc{\cn}{\C^n}
\nc{\rn}{\R^n}

\nc{\SO}{SO}
\nc{\Sp}{Sp}
\nc{\SU}{SU}

\nc{\Wtree}{W_{\mathrm tree}}
\nc{\Weff}{W_{\mathrm eff}}

\def\complex      {{\mathbb C}}

\def\projective   {{\mathbb P}}

\def\reals        {{\mathbb R}}
\def\zet          {{\mathbb Z}}

\def\be		{\begin{equation}}
\def\ende		{\end{equation}}

\def\revise#1       {\marginpar{\rule{2mm}{1cm} #1}}

\def\CC{\complex}
\def\ZZ{\zet}
\def\RR{\reals}
\def\PP{\projective}
\def\RP{\RR\PP}
\def\N{{\cal N}}

\def\bea		{\begin{eqnarray}}
\def\eea		{\end{eqnarray}}
\nc{\e}{{\rm exp}}

\nc{\cosech}{{\rm cosech}}

\nc{\Li}{{\rm Li_{2}}}

\nc{\li}{\lambda_{i}}
\nc{\lj}{\lambda_{j}}
\nc{\lk}{\lambda_{k}}
\nc{\laml}{\lambda_{l}}

\nc{\mi}{\mu_{i}}
\nc{\mj}{\mu_{j}}
\nc{\mk}{\mu_{k}}
\nc{\ml}{\mu_{l}}

\nc{\om}{\omega}

\nc{\ra}{\rightarrow}

\nc{\non}{\nonumber}

\begin{document}

\title{Inherited Duality and Quiver Gauge Theory}

\date{June 2004}

\author{Nick Halmagyi, Christian R\"omelsberger and Nicholas P.  
Warner\\[0.4cm]
\it Department of Physics and Astronomy\\
\it University of Southern California \\
\it Los Angeles, CA 90089, USA \\[0.2cm]
}

\Abstract{
We study the duality group of $\widehat{A}_{n-1}$ quiver gauge
theories, primarily using their M5-brane construction.
For $\CN=2$ supersymmetry, this duality group was first noted by Witten
to be the mapping class group of a torus with $n$ punctures. We find  
that it is a
certain quotient of this group that acts faithfully on gauge couplings.
This quotient group contains the affine Weyl group of  
$\widehat{A}_{n-1}$, $\ZZ_n$ and
$SL(2,\ZZ)$. In fact there are $n$ non-commuting $SL(2,\ZZ)$ subgroups,  
related
to each other by conjugation using the $\ZZ_n$.
When supersymmetry is broken to $\CN=1$ by masses for the adjoint  
chiral superfields,
an RG flow ensues which is believed to terminate at a CFT in the  
infrared.
We find the explicit action of this duality group for small values of  
the
adjoint masses, paying special attention to when the sum of the masses is  
non-zero.
In the $\CN=1$ CFT, Seiberg duality acts non-trivially
on both gauge couplings and superpotential couplings and we interpret  
this duality as inherited
from the $\CN=2$ parent theory.  We conjecture the action of S-duality
in the CFT based on our results for small mass deformations.
We also consider non-conformal deformations of these $\CN=1$ theories.   
The
cascading RG flows that ensue are a one-parameter generalization of  
those found
by Klebanov and Strassler and by Cachazo {\it et. al.}.  The  
universality exhibited
by these flows is shown to be a simple consequence of paths generated by
the action of the affine Weyl group.
}

\enlargethispage{1.5cm}

\makepapertitle

\vfill \eject

\body

\version\versionno


\section{Introduction}

For nearly a decade, branes have provided crucial insights into
supersymmetric gauge theories. Two arenas where branes have
found particular success are Maldacena's gauge/gravity duality  
\cite{Maldacena:1997re}
and IIA suspended brane constructions
\cite{Klemm:1996bj,Hanany:1996ie, Witten:1997sc}. 
The duality has proved to be a useful arena
for analyzing renormalization group (RG) flows, in many cases
exact supergravity solutions can be found \cite{Freedman:1999gp,  
Pilch:2000fu,
Klebanov:2000hb}.
In ten dimensions these flows are driven by non-trivial profiles
for the $p$-form fluxes and/or geometric moduli. The IIA constructions
on the other hand, have proved perhaps a little less useful, at least
in terms of RG flows but in this paper we will find great utility for  
these
setups in uncovering explicit actions of duality groups.

With the goal of studying duality symmetries within a certain class of  
RG flows,
we study  the ${\cal N}=2$ quiver gauge theory that is dual to IIB  
string theory
on $AdS_{5}\times S^{5}/\ZZ_n$ \cite{Douglas:1996sw, Kachru:1998ys}.
Five-dimensional, ${\cal N}=8$ gauged supergravity has proven very  
useful
in the analysis of holographic flows in ${\cal N}=4$ Yang-Mills. It was
argued in \cite{Corrado:2002wx} that ${\cal N}=4$ gauged supergravity,
with very specific couplings to ${\cal N}=4$ vector and tensor  
multiplets
would provide  a similar tool for ${\cal N}=2$ quiver gauge theories.  
The
most general such supergravity theory was constructed in  
\cite{Dall'Agata:2001vb}, and
the precise version relevant to quiver gauge theories was found
and analyzed in \cite{Corrado:2002wx}.  It is believed (but not proven)
that this theory is a consistent truncation of the ten-dimensional  
theory, and so
results derived in five dimensions should precisely correspond to  
ten-dimensional
backgrounds. The results of \cite{Corrado:2002wx} thereby made some
intriguing predictions for IIB supergravity backgrounds.  In  
particular,  it was shown that,
after gauging, there is an $SU(1,n)$ global symmetry group that remains  
unbroken.
The $SU(1,n)$ acts on scalars holographically dual to the complex gauge  
couplings
of each  gauge group, implying that each gauge coupling can be set
to an arbitrary value. (In fact the $SU(1,n)$ acts on more scalars
than just those dual to gauge couplings. This will be described in more  
detail
below.)  In the untwisted IIB string theory on $AdS_5\times S^5$, this
duality group is simply the $SU(1,1) \equiv SL(2,\IR)$ which is
broken to $SL(2,\ZZ)$ in string theory, and hence in the finite N gauge  
theory.
Thus one must expect, in general, that the $SU(1,n)$ will be broken to a
discrete subgroup for finite-rank  gauge groups.    This led us
to the present investigation of duality symmetries from the perspective
of field theory.

There is another description of these ${\cal N}=2$ quiver gauge theories
\cite{Klemm:1996bj,  Witten:1997sc} that is T-dual to the IIB  
description.
This construction is the classic IIA suspended brane construction  
lifted to M-theory.
In this construction, the D4-branes and the NS5-branes
that they are suspended between, lift to M-theory as a single M5-brane  
wrapped
on the Seiberg-Witten curve. For our purposes, the advantage of
this M-theory picture over its IIB counterpart is that the duality
symmetries can be easily and precisely described without the need to  
take a
't Hooft limit.  In \cite{Witten:1997sc}   Witten noted that the  
duality group that acts
on complexified gauge
couplings is the mapping class group of a torus with $n$ punctures,  
which following the
notation used in \cite{Birman74} we denote $M(1,n)$.

The group $M(1,n)$ has been much studied by mathematicians and is known  
to be an
extension of the torus braid group on $n$-strands by $SL(2,\ZZ)$  
\cite{Birman74}.
We find that the representation of this group
which acts on gauge couplings is not faithful but it fails to be  
faithful in a very simple
way. If one supplements $M(1,n)$ with the condition that permutations  
are involutions
instead of braiding, i.e. $s_{i}^2=1$ where $s_i$
exchanges the $i$-th and $i+1$-th punctures, then one does obtain a  
faithful representation.
With these extra relations this new group has the following subgroups:  
the
$\widehat{A}_{n-1}$ affine Weyl group,
a $\ZZ_n$ which rotates the nodes of the quiver and of course  
$SL(2,\ZZ)$.
It has already been suggested that affine quiver theories have the  
affine Weyl
group as a duality group \cite{Cachazo:2001sg}, it is nice to find a  
direct
connection to the M-theory picture.
We find that the $SL(2,\ZZ)$ has a preferred node and thus there
are $n$ copies of $SL(2,\ZZ)$ related by conjugation by elements of  
$\ZZ_n$.
These three subgroups appear in an identical fashion in the
context of $SU(n)$ WZW models, this is a tantalizing observation but we  
are unable to provide
a deeper connection.

These field theories admit a family of relevant deformations by mass
terms for adjoint chiral superfields,
which according to the methods of Leigh and Strassler  
\cite{Leigh:1995ep} should flow
to a non-trivial CFT in the IR \cite{Corrado:2002wx, Corrado:2004bz}.
The main purpose of this paper is to study the duality group of these  
flows.
For the $\ZZ_n$ orbifold there are thus $n$ independent mass parameters.
One of these mass parameters is the sum of the masses on the nodes
of the quiver and, being invariant under permutations, it comes from the
untwisted sector of the gauge theory. Such a deformation of the gauge  
theory
is therefore is dual to the $\ZZ_n$ orbifold of the Pilch-Warner flow  
\cite{Pilch:2000fu}.
The other mass parameters are dual to complexified K\"ahler moduli
of the blow-ups of the orbifold singularities \cite{Corrado:2002wx}.
For  $n=2$,  the IR fixed point of the resulting flow is dual to
$AdS_5\times T_{1,1}$ \cite{Klebanov:1998hh}, and the twisted-sector  
flows for
the general orbifolds were described in \cite{Gubser:1998ia}.
One may also consider a relevant deformation by a combination of
masses in both  the twisted and untwisted sectors, and these too should  
flow
to a CFT in the IR \cite{Corrado:2002wx} however the
ten-dimensional supergravity dual for these IR
fixed points, let alone the entire flows has yet to be constructed.

On the other hand, while the potentials for five-dimensional gauged  
supergravity
theories tend to be unwieldy,  the analysis of special sub-sectors
can prove relatively simple. For flows described above it was easy to
show \cite{Corrado:2002wx} that  the $SU(1,n)$ invariance of the theory  
also
acts on the scalars dual to adjoint masses.  In particular, that the  
compact,
$SU(n)$ subgroup acts on them in the fundamental representation.
The holographic dual of the flow involving only the untwisted sector
mass is explicitly known, both in five dimensions  
\cite{Freedman:1999gp},
and in ten-dimensions \cite{Pilch:2000fu}.  One can thus use the
$SU(n)$ symmetry to map the known FGPW to any flow involving any  
combination
of twisted and untwisted sector masses.  This symmetry is manifest in  
five dimensions,
but is far from obvious in ten dimensions:  The untwisted sector mass is
dual to a topologically trivial $B$-field flux, while the twisted  
sector masses
involve K\"ahler moduli of blow-ups. Such an unusual duality rotation
was one of the more intriguing geometric predictions of  
\cite{Corrado:2002wx}.

In the IIA picture, masses for the adjoint fields
are introduced by rotating the NS5-branes \cite{Barbon:1997zu}
and explicit formulas in terms of the
ten-dimensional geometry have been obtained when the masses are small.
Using this, we are able to describe the duality symmetries of
these mass deformations and again we find the affine Weyl group and  
$\ZZ_n$,
however, only the $S$-transformation from $SL(2,\ZZ)$ acts  
non-trivially.
Each mass behaves under a Weyl reflection as a Dynkin label and the sum  
of
the masses, the `global mass', is the level and is thus invariant.

As mentioned above, the RG flows that ensue from these mass  
deformations should
all go to non-trivial, conformal fixed points.  Indeed, the holographic  
solution
  \cite{Corrado:2002wx} suggests that the fixed line of \cite{  
Leigh:1995ep} becomes a family
of such lines parameterized by $\IC \IP^{n-1} = SU(n) /(SU(n-1) \times  
U(1))$.
By integrating out the massive fields we  get a quartic superpotential
and a new set of superpotential couplings, which are proportional to  
the inverse
of the masses. In the suspended brane construction, Seiberg duality  
corresponds
to interchanging  NS5-branes \cite{Elitzur:1997fh}.
In \cite{Corrado:2004bz} it was shown that Seiberg duality
acts non-trivially on the superpotential couplings and a 1-parameter
generalization of the Klebanov-Strassler cascading RG flow to include
non-trivial profiles for the $3$-form flux was conjectured. These
new flows were shown to exhibit universality in the sense that
after a large number of Seiberg dualities, the sum of the inverse of  
the superpotential
couplings approaches zero, which is the condition for vanishing  
$3$-form flux.
In this paper we point out that the inverses of superpotential couplings
transform under Seiberg duality as Dynkin labels under an affine Weyl  
reflection.
The universality is then a simple statement about paths generated by  
the affine Weyl group.
We interpret this duality as `inherited' from the parent ${\cal N}=2$  
theory.

Inherited dualities are an interesting phenomenon \cite{strassler}.
The concept first appeared as an attempt to prove that in special cases,
namely when $N_f=2N_c$,  Seiberg duality is inherited from ${\cal N}=2$
$S$-duality \cite{Argyres:1996eh, Leigh:1995ep}. Since then,
$S$-duality of the $N=1^{*}$ flow has been intensively studied
  \cite{Dorey:2000fc, Dorey:2001qj}. The  authors of \cite{Dorey:2001qj}
  study $\CN=2$ quiver gauge theories with nonzero masses for the  
hypermultiplets
as well for the adjoint chiral superfields, these theories flow to  
either confining or Higgs vacua.
Inherited duality for ${\cal N}=4$ SYM deformed by a single mass term  
has also been studied
\cite{Argyres:1999xu}. In this paper we conjecture that the ${\cal N}=1$
$\widehat{A}_{n-1}$ quiver gauge theories have $n$ $SL(2,\ZZ)$ duality  
symmetries
distinct from Seiberg duality. These $SL(2,\ZZ)$  groups do not commute  
and thus
there is no sense in which we have a `diagonal' or `overall'  
$SL(2,\ZZ)$. We establish that
all these symmetries are inherited from their ${\cal N}=2$ parent.


\section{The $\N=2$ Duality Group}

In \cite{Witten:1997sc}, Witten constructed four dimensional ${\cal  
N}=2$,
$\widehat{A}_{n-1}$ quiver gauge theories by lifting a IIA brane  
construction
to M-theory. The starting point is IIA with $n$ separated NS5-branes and
$N$ D4-branes suspended between each of them. Specifically, the  
asymptotic
coordinates of the $i$-th NS5-brane are $x^{7}=x^{8}=x^{9}=0$ and  
$x^{6}=x^{6}_{i}$.
The D4-branes have four world volume coordinates in common with the NS5-
branes $(x^{0},\ldots,x^{3})$, they also have finite extent in the
$x^{6}$ direction thus the gauge theory living on their worldvolume is
effectively four dimensional. There are $n$ gauge coupling constants,
given by
\be
\frac{1}{g_{i}^2}=\frac{x^{6}_{i+1}-x^{6}_{i}}{8\pi g_s L}\,.
\label{gcouplings}
\ende
The affine quiver gauge theories are engineered when the $x^{6}$  
direction
is periodic\footnote{These are referred to as the {\it elliptic models}  
in
\cite{Witten:1997sc}}, with period, $L$.   There are then $n$ NS-5  
branes, and the indices
  in (\ref{gcouplings}) are taken cyclically, except that in going from
$n^{\rm th}$ brane to the first brane, one continues around
the circle, and picks up a full period so that
\be
\sum_{i=1}^n \, \frac{1}{g_{i}^2} ~=~ \frac{1}{8\pi g_s}\,.
\label{coupsum}
\ende
Upon lifting to M-theory, the NS-5 branes will be
separated in the $x^{10}$ direction as well and this separation is
interpreted as the theta angle, namely
\be
\theta_{i}=\frac{x^{10}_{i+1}-x^{10}_{i}}{R}\,.
\ende
The periodicity properties are now represented in terms of a
torus, $E_{\tau}$, is constructed from the $x^{6}$ and $x^{10}$  
directions.
First, one must impose that $x^{10}$ is periodic:
\be
x^{10}\sim x^{10}+2\pi R,,
\label{tenperiod}
\ende
which guarantees the standard periodicity under $\theta_j \to
\theta_j + 2\pi$ in each gauge theory factor separately.  The  
periodicity in
the $x^6$ direction is now modified to:
\be
   ( x^{6}\,, x^{10} )\sim ( x^{6}+2 \pi L \,, x^{10}+\theta R )
   \label{sixperiod}
\ende
If one now defines:
\bea
p_{i} &~\equiv~&   \frac{i\, x^{6}_{i}}{16\pi^2 g_s L} ~+~
\frac{ x^{10}_{i}}{2 \pi R}\,, \\
\tau &~\equiv~&  \frac{i}{8\pi g_s } ~+~
\frac{\theta}{2\pi}\,,
\eea
then the $p_i$ are points on the elliptic curve, $E_{\tau}$, with  
Teichmuller
parameter, $\tau$.   Individual gauge couplings are then
differences between points on this torus, and the cyclically defined
sum, (\ref{coupsum}), becomes
\be
\sum_{i=1}^n \,   (p_{i+1}  - p_{i } ) ~=~ \tau\,.
\label{cyclicsum}
\ende
Specifically, because the $D4$ branes wrap the torus along the
$x^6$-direction in the quiver gauge theory, we thus take:
\be
p_{n+1}  =  p_{1 }  +  \tau\,, \qquad p_{0}  =  p_{n }  -  \tau\,,
\label{cyclicpoints}
\ende
We will discuss this below.

Witten thus concludes that the moduli space of coupling constants of the
quiver gauge theory is the moduli space ${\cal M}_{1,n}$ of a 2-torus
with $n$ distinct, unordered punctures. The duality group of the four
dimensional gauge theory is then the mapping class group $M(1,n)$.


\subsection{Generators and Relations}

We now describe explicitly how this mapping class group acts on gauge
couplings.  The covering space of the moduli space ${\cal M}_{1,n}$
can then be parametrized by $\tau$ and the $n$ marked points $p_{i}$ in  
the
complex plane. Since only differences of the $p_i$ have physical  
meaning, these $n$
points are only defined up to an overall shift:
\be
(\tau,p_1,\ldots,p_n)\sim(\tau,p_1+c,\ldots,p_n+c)\,,
\ende
for any complex constant, $c$.
We shall call this basis for the representation of our group the
{\it point basis}. Then the generators of $M(1,n)$ are the usual  
generators
$S$ and $T$ of $M(1,1)\sim SL(2,\ZZ)$, combined with permutations
of the marked points, $s_i$, and individual lattice shifts $T_i$ and  
$t_i$.
The explicit action of these generators are:
\bea
&& S:\ (\tau,p_1,\ldots ,p_{n})\ra
(-\frac{1}{\tau},\frac{p_1}{\tau},\ldots ,\frac{p_n}{\tau}), \non \\
&& T:\ (\tau,p_1,\ldots ,p_{n})\ra (\tau + 1,p_1,\ldots ,p_{n}), \non \\
&& T_{i}:\ (\tau,p_1,\ldots ,p_{n})\ra
(\tau,p_1,\ldots, p_i+1 ,\ldots,p_{n}), \non \\
&& t_{i}:\ (\tau,p_1,\ldots ,p_{n})\ra
(\tau,p_1,\ldots, p_i+\tau ,\ldots,p_{n}), \non \\
&& s_{i}:\ (\tau,p_1,\ldots,p_{i},p_{i+1},\ldots ,p_{n})\ra
(\tau,p_1,\ldots,p_{i+1},p_{i},\ldots ,p_{n}),\ i\neq n, \non \\
&& s_{n}:\ (\tau,p_1,\ldots ,p_{n})\ra (\tau,p_n-\tau,p_2\ldots  
,p_{n-1},p_{1}+\tau).\non
\eea
The action of $s_n$ differs in form from the other $s_i$ because of the
cyclic structure of the $p_j$ arising from the wrapping of the $D4$  
branes.
Indeed, one can obtain this formula if one properly applies
the convention (\ref{cyclicpoints}).   This rule will also emerge
from the discussion of gauge couplings, where we require that
the permutations  act in the same manner upon all gauge couplings.

There are non-trivial relations between these generators and some of the
generators are actually redundant. First, note that $S$ and $T$ are the  
usual
generators of $SL(2,\ZZ)$ with $S^2$ acting non-trivially on the  
$n$-points
but trivially on $\tau$. Less obvious are the relations obeyed by the  
other generators.
To elucidate  these we first introduce another basis for the  
representation called the
{\it gauge coupling basis}. This basis is $(\tau_1,\ldots , \tau_n)$  
where
\be \label{gauge}
\begin{array}{l}
\tau_i = p_{i+1}-p_{i},\ i\neq n,  \\
\tau_n=\tau+p_1-p_n.
\end{array}
\ende
The overall gauge coupling constant $\tau$ can be recovered as
$\tau=\sum_i\tau_i$. In this basis the modular group acts as follows
\footnote{For $n=2$ the action of the $s_i$ is slightly different, as  
is familiar
from the action of affine Weyl groups on fundamental weights.}
\bea
&& S:\ (\tau_1,\ldots,\tau_n)\ra
\left(\frac{\tau_1}{\tau},\ldots,
\frac{\tau_{n-1}}{\tau},
\frac{\tau_n}{\tau}-1-\frac{1}{\tau}\right),\non \\
&& T:\ (\tau_1,\ldots,\tau_n)\ra  
(\tau_1,\ldots,\tau_{n-1},\tau_n+1),\non \\
&& T_{i}:\ (\tau_1,\ldots ,\tau_{n})\ra
(\tau_1,\ldots,\tau_{i-1}+1, \tau_i-1,\ldots,\tau_{n}), \non \\
&& t_{i}:\ (\tau_1,\ldots ,\tau_{n})\ra
(\tau_1,\ldots,\tau_{i-1}+\tau, \tau_i-\tau ,\ldots,\tau_{n}), \non \\
&& s_{i}:\ (\tau_1,\ldots,\tau_{i-1},\tau_{i},\tau_{i+1},\ldots  
,\tau_{n})\ra
(\tau_1,\ldots,\tau_{i-1}+\tau_{i},-\tau_{i},\tau_{i+1}+\tau_{i},\ldots  
,
\tau_{n})\, \non
\eea
where we now interpret these formula with the exact cyclic convention:
$\tau_{n+1} = \tau_{1}$ and $\tau_{0} = \tau_{n}$.  This fixes the
form of $s_n$ above.

We can now find relations between these generators.
Firstly, the $T_i$ and $t_i$ are related
by the $S$-transformation
\be
t_i=S^{-1}T_i S\quad{\rm and}\quad T_i^{-1}=S^{-1}t_iS.
\ende
We introduce $\omega$, the generator of $\ZZ_n$, which in the gauge
coupling basis acts as $\tau_i \ra \tau_{i+1}\,, \forall \ i$ and can be
easily seen to satisfy
\be
\omega= t_{i} s_{i-1} s_{i-2}\ldots s_{i+1}.
\ende
We can build $T_i$ out of $T$ and the cyclic permutations $\omega$
\be
T_i=\omega^{n-i+1}T\omega^{-1}T^{-1}\omega^{-n+i}.
\ende
Finally, the reflection operations $s_i$ are related by cyclic
permutations
\be
s_{i+1}=\omega^{-1}s_i\omega.
\ende
Thus we have shown that a minimal set of generators is
$\{S,T,s_1,\omega \}$ although for a nice presentation
we include $s_i,t_i,T_i$ and exclude $\omega$, see the appendix for
such a presentation.

It is quite interesting that the $SL(2,\ZZ)$ subgroup described
above chooses a preferred node of the quiver. The $\ZZ_n$ subgroup
rotates the nodes of the quiver and thus there are actually $n$  
different
$SL(2,\ZZ)$ subgroups, all related by conjugation by an
element of $\ZZ_n$. Note that the different $SL(2,\ZZ)$
subgroups do not commute with each other. This means
that there is no diagonal subgroup which could be the `overall'
$SL(2,\ZZ)$.

 From the point of view of holographic RG flows in type IIB string theory
this might seem a little strange at first. In type IIB theory there
is only one $SL(2,\ZZ)$ duality group in ten dimensions. However
in a compactification of type IIB theory on an orbifold it is not
clear how the $SL(2,\ZZ)$ acts on the twisted sector fields. In
order to determine that, one has to blow up the singularity, such
that the whole geometry is smooth. Then the action of $SL(2,\ZZ)$ is
well defined. The stringy geometry provides $n$ different ways
to blow up the singularity. These blowups are related by the $\ZZ_n$
quantum symmetry as we have discovered above. As expected, each  
$SL(2,\ZZ)$ subgroup acts in the same way on the untwisted fields.

For a generic set of gauge couplings, it can happen that after
one of the foregoing duality transformations one or more of the  
individual couplings have negative imaginary part. This is unphysical. So one
must use these duality transformations with care. For example,
after an $S$ transformation, the strength of the gauge couplings
is determined by the differences of positions of NS5-branes in the
$x^{6}$ direction. One can then use the involutions, $s_i$,
to reorder the NS5-branes such that the $n$-tuple
$(x^{6}_1,\ldots,x^{6}_n)$ is restored to increasing
order.   As we will see in the next section, the $s_i$ are, in fact,  
Seiberg dualities
acting on the $i^{\rm th}$ node  of the quiver theory.  This is  
therefore in perfect accord
with the previous observations  \cite{Cachazo:2001sg}
that Seiberg duality can be used to prevent $1/g_{i}^2$ from
becoming negative.

The mapping class group $M(1,n)$ has been studied in
detail by Birman \cite{Birman69, Birman74}. This group is an extension  
of the torus braid
group $B_n(T^2)$ by $SL(2,\ZZ)$. From the presentation of $B_n(T^2)$
given in \cite{Birman69} we can see that our representation is not  
faithful.
If we supplement $B_n(T^2)$ with the relations $s_{i}^2=1$ we
recover the presentation provided in the appendix. The non-trivial  
homeomorphism
$s_i^2$ of $M(1,n)$ has become trivial, in the IIA picture
this is due to brane-antibrane annihilation.


\section{The $\CN=1$ Duality Group}

We now study deformations of the $\CN=2$ quiver gauge theory by masses  
for the adjoint
scalars. The methods of Leigh and Strassler \cite{Leigh:1995ep} predict  
that,
assuming the hypermultiplet
masses are set to zero, the end-point of the resulting RG flow will
be a non-trivial CFT \cite{Corrado:2002wx, Corrado:2004bz}.


\subsection{Small Mass Deformations}

To understand the ${\cal N}=1$ duality group we will again use the
5-brane construction in M-theory. The construction of $\CN=2$
gauge theories requires that the M5-brane wraps a complex curve
that is embedded in the two complex dimensions with holomorphic  
coordinates
$u=x^4+ix^5$ and
$w=x^6+ix^{10}$.  More generally, a $\CN=1$ theory merely requires that  
the M5-brane
wrap a curve embedded in three complex dimensions, with the extra  
dimension
given by $v=x^7+ix^8$.    In the IIA theory, the soft breaking of the  
$\CN=2$ theory
down to $\CN=1$ supersymmetry can be achieved by essentially tilting  
the NS-5 branes
relative to  one-another.  In M-theory, we need the  M5-brane to wrap
a curve in $(u,v,w)$-space  and for large $(u,v,w)$ the NS-5 branes   
must be
asymptotic to  points on the elliptic curve, $E_{\tau}$, defined by
(\ref{tenperiod}), (\ref{sixperiod}) in the $w$-direction and by   
straight lines in the
$(u, v)$-direction.  The directions of these lines  can be   
parametrized by points, $z_i$,
on a $\PP^1$ that has $u$ and $v$ as homogeneous coordinates.
Therefore, the asymptotic M5-brane data is
the collections of points, $p_i$, on the torus and a set of points,  
$z_i$
on a $\PP^1$.

As before, the gauge coupling constants $\tau_i$ are given
by (\ref{gauge}). When at least
two of the NS5-branes, or equivalently the asymptotic legs of the  
M5-brane, fail to be
parallel,  ($z_i \not = z_j$ for some $i$ and $j$)  the $\CN=2$  
supersymmetry is softly broken
by small masses $m_i$ for the adjoint scalars \cite{Barbon:1997zu}.
There is, however, a  problem:  Given a quiver theory with $n$ nodes,
there must be $n$ independent mass parameters for the $\CN=1$  chiral  
multiplets
in the $\CN=2$  vector multiplets.  However, a set of $n$ NS-5 branes  
has only
$(n-1)$ independent relative angles between them, and so the tilting of  
branes
describes all but one of the mass parameters.  We will resolve this  
issue
momentarily.

For a generic angle between the two NS5-branes, there is also an  
ambiguity
in determining the adjoint scalar. If the branes are almost parallel,  
the
canonical choice for the scalar is the position of the suspended  
D4-brane along
the direction of the NS5-branes. However, there is a second, very heavy  
scalar
for the motion of the suspended D4-branes perpendicular to the  
direction of
the NS5-branes. For this reason, a mass formula valid for
arbitrary masses should be a function on a double cover of the $\PP^1$.
The field theoretic meaning of the additional adjoint scalar which must
be integrated in along the flow is unclear. At any rate, we will
work primarily with with the small mass formula.  This means that all
the branes are very nearly parallel to each other asymptotically.

The group acting in the $(u,v)$-directions that preserves the complex
structure is $GL(2,\CC)$. In order to preserve the holomorphic 3-form,
this rotation goup has to be restricted to $SL(2,\CC)$.  Since the  
branes
are all almost parallel, we can use the  $SL(2,\CC)$ to rotate them so  
that
they are all nearly asymptotic to the $v$-direction.  This means that
they can all be  represented small values of  the inhomogenous
coordinates, $z_i = \frac{u_i}{v_i}$,
on $\PP^1$.   The mass parameters, $m_i$, in the superpotential for the
adjoint scalar $\Phi_i$ has to be a holomorphic function of $z_i$ and
$z_{i+1}$. Further, the mass must vanish iff the two 5-branes are
parallel. From this we get the following approximation when the masses
are small:
\be
m_i=z_{i+1}-z_i.
\ende
Observe that the sum of the masses vanishes, and so, at least for small  
$m_i$,
the missing mass parameter is the total, or global, mass parameter,
$m=\sum_{i=1}^{n}m_i$.

In \cite{Witten:1997sc} a similar problem was encountered and solved
for the independent hypermultiplet mass terms in the $\CN=2$ theory.
The solution was to fiber the mass parameters in a non-trivial manner
over the torus:  Traveling around one the cycles of the torus
resulted in a non-trivial (affine) translation of the masses.  This
affine shift, or connection on the fibration thus   provided the  
non-trivial
total mass.     Here we use a similar solution, but we cannot use
an affine fibration of the NS-5 branes as this would offset the
D4-branes suspended between them, and thus generate unwanted
hypermultiplet masses.

To generate a non-zero global mass  we should fiber $\CC^2$
described by $(u,v)$ over $E_{\tau}$, and only use the
$SL(2,\CC)$ structure group, so as preserve the supersymmetry
and fix the origin in $\CC^2$ so as to avoid displacing
the D4 branes.      Since the global mass can be turned off  
continuously,
the fibration has to be topologically trivial.
The moduli space of such a fibration is classified by an embedding of
$\pi_1(E_{\tau})=\ZZ\oplus \ZZ$ into the maximal
torus\footnote{Note that the maximal torus of $SL(2,\CC)$ is $\CC^*$.}  
of $SL(2,\CC)$.
If one considers  the projective action of $SL(2,\CC)$ on the $\PP^1$,
then the obvious choice of maximal torus is to take the complexified
rotations (rotations and scale transformations) $ z \to  a \, z $  that  
fix the origin
and the point at infinity.  However, this does not move branes located
at the origin, and so we conjugate this torus so that the fixed points  
are
$z=1$ and $z=-1$   on the $\PP^1$.   That is, we take the torus to be
$z \to \frac{a z+ b}{b z +a}$, with $a^2 - b^2 =1$.
The effect of such transformations on points near $z=0$ is
an affine shift: $z \to z+\frac{b}{ a}$, and thus if one travels
around the torus, one generates an  overall shift in the
global mass parameter.

We also see that for our purposes, we can approximate
the $\PP^1$ bundle by an affine fibration of $\CC$ over $E_{\tau}$,
which is incidentally also a principal $\CC$ bundle.  Our construction  
therefore
closely parallels the  construction of a hypermultiplet global mass via
an affine  fibration in \cite{Witten:1997sc}.

A principal $\CC$ bundle over $E_{\tau}= T^2$ has two gluing functions,  
one for each
cycle of the base, which are encoded in
\bea
&& (p_i,z_i)\sim (p_i+1,z_i+a), \\
&& (p_i,z_i)\sim (p_i+\tau,z_i+b).
\eea
However, there is a gauge freedom that means that only one
such parameter is physical.  In general the gauge transformation means
that we can shift the fiber coordinate by an arbitrary holomorphic  
function of the base
variable:   $z_i\ra z_i'=z_i+f(p_i)$.   However in order to remain  
consistent
with the constant gluing functions above, $f(p_i)$ must be a linear  
function.
Further, the constant part of this linear function
has no effect on the gluing functions and can be set to zero.
Thus we have a gauge freedom: $g: (p,z) \to (p,z +c\, p)$ for some  
constant,
$c$.  This gauge transformation also modifies the gluing functions, and  
so
we have:
\bea
(\tau,a,b,p_i,z_i)&\stackrel{g}{\ra}& (\tau,-c\cdot 0+a+c\cdot1,
-c\cdot0+b+c\cdot\tau,p_i,z_i+cp_i) \non \\
&=&(\tau,a+c,b+c\tau,p_i,z_i+cp_i) \non
\eea
By choosing $a=-c$ we get a trivial gluing function associated to the
$\theta$-cycle of $T^2$, while the remaining gluing function is  
identified
with the mass:
\bea
&& m_i=z_{i+1}-z_i,\\
&& m_n=z_1-z_n+m \,.
\eea
This periodicity now precisely parallels that of the gauge coupling  
constants.
The gauge choice above (associating $m$ with the $x^6$ cycle)
ensures, that $T_i$ acts as the
identity on mass parameters as required by physics.

 From the foregoing definition it is immediate that the masses transform
in the same way as the gauge coupling constants under the permutation
symmetries which generate the affine Weyl group (now with $\sum m_i=m$
as the level) when acting on the masses.
In order to understand the action of $S$-duality on the mass  
parameters, we will need
to explicitly work out gauge transformations in local coordinates.

The action of $S$-duality violates the gauge choice made above, and
to restore the gauge it must be followed by a gauge transformation
with $c=m$.  The geometrical data thus transforms as:
\bea
(\tau,0,m,p_i,z_i)&\stackrel{S}{\ra}&  
(-\frac{1}{\tau},m,0,\frac{p_i}{\tau},z_i)\\
&\stackrel{g}{\ra}&  
(-\frac{1}{\tau},0,\frac{m}{\tau},\frac{p_i}{\tau},z_i- 
m\frac{p_i}{\tau}).
\eea
 From now on we will leave implicit the fact that $S$-duality is coupled  
with
the foregoing gauge transformation.
It is instructive to check that $S^{2}$ acts in the following way
\be
(\tau,0,m,p_i,z_i)\stackrel{S^2}{\ra}(\tau,0,-m,-p_i,z_i).
\ende
We now have the complete action of the duality group on the masses,  
namely
\bea
&& S:\ (m,m_1,\ldots,m_n)\ra
\left(\frac{m}{\tau},m_1-\frac{m\tau_1}{\tau},\ldots,
m_{n-1}-\frac{m\tau_{n-1}}{\tau},
m_n-\frac{m \tau_{n}}{\tau}+\frac{m}{\tau}\right),\non \\
&& t_i:\ (m,m_1,\ldots,m_{i},\ldots ,m_{n}) \ra  
(m,m_1,\ldots,m_{i-1}+m, m_i-m,\ldots,m_n), \non \\
&& s_{i}:\ (m,m_1,\ldots,m_{i-1},m_{i},m_{i+1},\ldots ,m_{n}) \non \\
&& \ \ \ \ \ra (m,m_1,\ldots,m_{i-1}+m_{i},-m_{i},m_{i+1}+m_{i},\ldots ,
m_{n}),\non \\
&& \omega:\ (m,m_1,\ldots, m_i,\ldots,m_{n-1},m_n) \ra(m,m_2,\ldots,  
m_{i+1},\ldots,m_n,m_1) \non
\eea
with the generators $\{T_i,T\}$ acting trivially.

We see from the foregoing construction that the $\CN=1$ duality group
is the same as the $\CN=2$ duality group. This is partly due to the fact
that the mapping class group of $\PP^1$ is trivial, so no new
symmetries are introduced in the $\CN=1$ theory.
There are still $n$ different $SL(2,\ZZ)$ subgroups of the duality  
group,
each related by conjugation by an element of $\ZZ_n$.


\subsection{Seiberg Duality}

For large mass deformations we must dispense with our geometric mass  
formula.
However, at a scale below the smallest mass scale, we can integrate out  
all the massive
fields and obtain a quartic superpotential for the hypermultiplet  
fields.
In the IR we can expect a nontrivial CFT and here Seiberg duality
acts non-trivially on couplings. As we will see, this action is in
agreement with what we derived for the small mass approximation, namely
they transform as Dynkin labels under an affine Weyl reflection and thus
we interpret this duality `inherited' from its ${\cal N}$$=2$ parent  
theory.

It has long been appreciated that quartic superpotentials play a special
role in Seiberg duality, namely the dual theory also has a quartic  
superpotential.
This idea was used to great effect in the celebrated
Klebanov and Strassler scenario \cite{Klebanov:2000hb}, which is  
included in our
analyis with $n=2$ and $m_1+m_2=0$. In the KS solution the  
superpotential
couplings are left invariant under Seiberg duality, this is a fixed  
point
of a symmetry group elucidated in \cite{Corrado:2004bz}. In general
the superpotential couplings are not invariant
under this operation.

We now review the action of Seiberg duality on the ${\cal N}=1$  
superpotential.
The $\N$$=2$ quiver has a unique renormalizable superpotential
\cite{Douglas:1996sw}:
\be
W_{{\cal N}= 2}=\sum_{i=1}^{n} \lambda_{i} \Phi_{i} \left( A_i B_i  
-A_{i-1} B_{i-1} \right),
\ende
where, for $\N$$=2$ supersymmetry $\lambda_i={\sqrt 2}$.
We deform by the relevant operator
\be
\Delta W = \sum_{i=1}^{n} \frac{m_i}{2} \Phi_{i}^{2}
\ende
which breaks supersymmetry to $\N$$=1$. Integrating out the massive  
adjoint
fields results in a  new superpotential

\be \label{superone}
W_{{\cal N}= 1}=\sum_{i=1}^{n} h_{i} \left( A_i B_i -A_{i-1} B_{i-1}  
\right)^2
\ende
where $h_i=-\frac{\lambda_{i}^{2}}{m_i}$.

There is a straightforward recipe for calculating the Seiberg dual  
theory
\cite{Seiberg:1994pq}. When one gauge group
becomes strongly coupled (relative to the groups of its adjacent nodes)  
one
introduces dual quarks and `fundamental' mesons. To dualize on the
$j^{\rm th}$ node one introduces a matrix of meson fields:
\be
M =
\begin{pmatrix}
A_{j-1} A_j  & A_{j-1} B_{j-1} \\ B_j A_j & B_j B_{j-1}
\end{pmatrix} \label{meson_matrix}
\ende
and the superpotential (\ref{superone}) becomes
\bea
W_{{\cal N}= 1}&=&\sum_{i=1}^{j-2} h_{i} \left( A_i B_i -A_{i-1}  
B_{i-1} \right)^2 \non \\
&&+ h_{j-1} \left( M_{12}^2 + (B_{j-2}A_{j-2})^2 - 2 M_{12}  
B_{j-2}A_{j-2} \right) \non \\
&&+ h_j \left( M_{12}^2 + M_{21}^2 - 2 M_{11} M_{22} \right) \non \\
&&+ h_{j+1} \left( M_{21}^2 + (A_{j+1} B_{j+1})^{2} - 2 M_{21} A_{j+1}  
B_{j+1}  \right) \non \\
&&+\sum_{i=j+2}^{n} h_{i} \left( A_i B_i -A_{i-1} B_{i-1} \right)^2  
\non \\
&& + y\left( M_{11} a_{j-1} a_{j} + M_{12} a_{j-1} b_{j-1} + M_{21}  
b_{} a_{j} +
M_{22} b_{j} b_{j-1} \right),,
\eea
where  we have also introduced `dual quarks', $  
(a_{j},a_{j-1},b_{j},b_{j-1})$.
Since the action is quadratic in these new fields, one can trivially  
integrate them out.
However, if the mesons, $M_{ij}$,  are massive then they can be  
integrated
out instead, and this results in a new quartic superpotential.
This has the effect of making the following
replacement  of the fields attached to the $j^{\rm th}$ node:
\be
(A_{j},A_{j-1},B_{j},B_{j-1})\rightarrow  
(a_{j},a_{j-1},b_{j},b_{j-1})\,.
\ende
However,  after integrating out the massive mesons the couplings
undergo non-trivial transformations.  If we  abuse  notation and
rescale the `dual quarks' according to:
\be \label{rescale}
\frac{y}{2 h_{j}}(a_{j},a_{j-1},b_{j},b_{j-1})\rightarrow  
(A_{j},A_{j-1},B_{j},B_{j-1})\,,
\ende
then the resulting superpotential is
\be
W_{{\cal N} = 1}= \sum_{i=1}^{n} h'_{i} \left( A_i B_i -A_{i-1} B_{i-1}  
\right)^2
\ende
where the couplings have transformed as follows,
\bea \label{weyl}
{h_{i}'}^{-1} &=& h_{i}^{-1},\ \ i\neq j-1,j,j+1 \non \\
  {h_{j-1}'}^{-1} &=& h_{j-1}^{-1}+ h_{j}^{-1} \non \\
   {h_{j}'}^{-1} &=& -h_{j}^{-1} \non \\
  {h_{j+1}'}^{-1} &=& h_{j+1}^{-1}+ h_{j}^{-1}.
\label{SeibDualAct}
\eea
Recalling from \cite{Corrado:2004bz} that the condition for a
generalized conifold is $\sum h_{i}^{-1}=0$, one can see that the
rescaling (\ref{rescale}) is necessary for the Seiberg dual
of a generalized confold to be a generalized conifold, for
which there is much evidence \cite{Cachazo:2001sg}. We see
from (\ref{weyl}) that $h_{j}^{-1}$ transform under Seiberg
duality as Dynkin labels of the $\widehat{A}_{n}$ Weyl group
transform under a reflection.

This is in agreement with the transformations that we derived
in the regime of small masses in the previous section.
We know that the action of $T_{i}$ and $T$ is a symmetry of
the $\CN=1$ field theory, they act on gauge couplings in an identical  
way
as for the $\CN=2$ theory and they leave the superpotential couplings  
invariant.
However it is desirable to provide a field theory derivation of the
$S$-transformation. Once this is shown the action of $t_i$ is immediate.
We leave this for future work.


\subsection{The Cascade}

So far we have considered theories with product gauge groups
with the rank of each group being equal. If one considers the $\CN=1$
theories but with ranks that are no longer all equal, then a nontrivial
RG flow ensues \cite{Klebanov:1998hh}.
This flow can be adequately described by a cascade of Seiberg
dualities on alternate nodes. In \cite{Cachazo:2001sg} it was
realized that the effect of this RG flow on gauge couplings can
be simply described by the motion of a `billiard ball' in the
fundamental Weyl chamber of the $A_{n-1}$ affine Weyl group.
When the ball bounces off a wall this corresponds to a Weyl
reflection or Seiberg duality. From the results of
\cite{Corrado:2004bz} we see that in addition, Seiberg
duality acts on superpotential couplings as a
fundamental affine Weyl reflection.
\begin{figure}[bth]
\centerline{ \epsfig{file=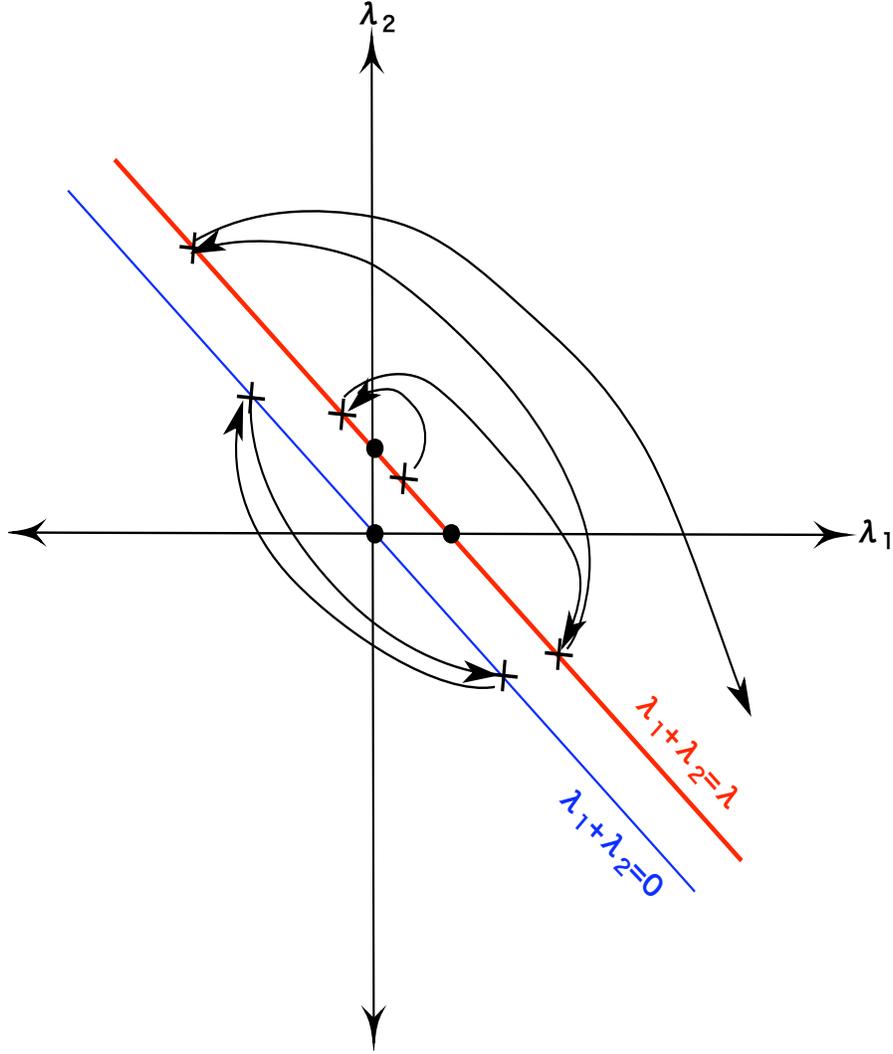,width=12cm,height=14cm}}
\caption{\sl Weyl reflections acting on fundamental weights of  
$\widehat{A}_1$.}
\label{weyling}
\end{figure}
One main result from \cite{Corrado:2004bz} was that there exists  
universality
in the IR of a general cascade associated to the $\widehat{A}_n$ quiver.
More precisely, define a UV theory to be one of the ${\cal N} =1$  
theories
discussed above but with the ranks of the gauge
groups unequal and an arbitrary initial set of superpotential couplings  
$h_i$.   Define
\be
h^{-1} ~\equiv~ \sum_{i=1}^{n}h_{i}^{-1}\, ,
\ende
and observe that this quantity is invariant under the actions of  
Seiberg dualities
(\ref{SeibDualAct}).  Consider the projective space with homogeneous  
coordinates
$[h_1^{-1},\ldots,h_n^{-1}]$, then a sequence of Seiberg dualities taken
around the quiver has the effect of translation along this
projective space.
There is a submanifold of the projective space defined by $h^{-1}=0$
and called  the `conifold subspace.'   Apart from being invariant under  
Seiberg duality,
this subspace is an attractor for all duality cascades   
\cite{Corrado:2004bz}.
Consider a theory  with  $h^{-1}\neq0$, then after a large number of  
such Seiberg dualities
a  subset of superpotential couplings will satisfy $|h_i^{-1}|\gg  
h^{-1}$.  As
explained in \cite{Corrado:2004bz},  the physical couplings are
the projective quantities, $h_i /h_j$, and since $h^{-1}$ is invariant,  
the physical
couplings are actually approaching the conifold subspace.

This property has a simple explanation in terms of paths in the affine  
Weyl group.
The couplings $h_{i}^{-1}$ are Dynkin labels for the reflection group,  
and
$h^{-1}=\sum_{i=1}^{n}h_{i}^{-1}$ is the level.
Consider the $\widehat{A}_1$ Weyl group with Dynkin labels  
$\lambda_1,\lambda_2$
and level $\lambda$. Under a Weyl reflection the level is left  
invariant, and
furthermore the pure translations in the affine Weyl group are  
proportional to the
level.   If we draw the line $\lambda_1+\lambda_2=\lambda$
in the plane as in Figure \ref{weyling}, then a Weyl reflection is a  
reflection in a hyperplane
which passes through either of the two points
$(0,\lambda)$ or $(\lambda,0)$ and is
perpendicular to this line. From the figure we see that
alternating reflections result in an arbitrary initial point (here we  
chose an initial point
to lie between $(\lambda,0)$ and $(0,\lambda)$)
being mapped toward infinity. In projective
space this point is moving toward the point $\lambda_1/\lambda_2=-1$.
As explained in \cite{Corrado:2004bz} it is the projective coordinates
$[h_1^{-1},h_2^{-1}]=[\lambda_1,\lambda_2]$ that are important  
parameters.
If the initial point was chosen to lie on the line with $\lambda=0$
then the two reflection planes coincide:  The translations collapse
because the level is zero.  This is  the well known fact that
only the subgroup of the affine Weyl group which is
isomorphic to the finite Weyl group acts faithfully on roots, and for   
$\widehat{A}_{1}$
this means that the point remains at $\lambda_1/\lambda_2=-1$.
This is a simple explanation for the universality found in  
\cite{Corrado:2004bz}.


\section{Discussion}

One interesting open problem that we have not addressed in this paper
is the relationship between the duality group found in the M-theory
construction and the $SU(1,n)$ found in five dimensional gauged  
supergravity.
The M-theory duality group is the symmetry group of a quiver gauge  
theory
with finite rank gauge group and we expect that it admits an embedding
into $SU(1,n)$ in the same sense that $SL(2,\ZZ)$ embeds into $SU(1,1)$.
There are certainly $n$ distinct, natural, non-commuting $SL(2,\ZZ)$  
subgroups
in $SU(1,n)$, and it is tempting to identify them with the subgroups
identified above.  One can also embed the $A_{n-1}$ affine Weyl group
into $SU(1,n)$. This is straightforward if one adds
another dimension to the weight spaces. This extra dimension is
the $L_0$ operator familiar from WZW models, but its is not clear how  
this relates
to the gauged supergravity.  Indeed, while parts of the duality group
identified here do fit into the $SU(1,n)$ of supergravity, we have not
yet succeeded in embedding the entire duality group in a manner  
consistent
with signatures of the appropriate invariant metrics.   In fact, this  
may not
be possible, since it is perhaps naive to hope that supergravity  
captures
all the dualities of the field theory.

There is a fascinating connection to WZW models which may not be  
unrelated.
The affine Weyl group and $\ZZ_n$ appear naturally in Kac-Moody  
algebras but further,
the parameters in the characters have modular transformations identical  
to
those found here for gauge couplings. Obviously a fundamental input
to defining a character is to have a well defined $L_0$ operator,  
something we are lacking
here.   Moreover, the familiar unitary representations of WZW models  
require an
integer level, whereas the level, $\tau$, for the field theory is an  
arbitrary complex number.

There is also the connection between WZW models, the mapping class  
group of
Riemann surfaces with punctures and Chern-Simons theory
\cite{Moore:1988uz, Moore:1989vd}. It would be
fascinating to find a connection between the F-terms of the  
four-dimensional
quiver gauge theory and Chern-Simons theory in three dimensions,  
perhaps along the lines
\cite{Witten:1992fb, Bershadsky:1993cx, Dijkgraaf:2002dh}.

\begin{acknowledgments}
This work is supported in part by funds provided by the DOE under grant  
number
DE-FG03-84ER-40168. The work of NH is supported in part by a Fletcher  
Jones Graduate
Fellowship from USC.
We would like to thank David Berenstein, Richard Corrado, Ruth Corran, Mike Douglas,
Chris Herzog, Timothy Hollowood,
Ken Intrilligator, Fyodor Malikov, Bob Penner, Johannes Walcher
and Ed Witten for useful discussions and correspondence.
\end{acknowledgments}

\begin{appendix}

\section{Presentation of the ${\cal N}=2$ Duality Group}

Here we write down a presentation for the ${\cal N}=2$ duality group.
The generators $\{s_i,t_j,T_j| i=1,\ldots,n-1\ j=1,\ldots,n\}$ are also  
generators
of $B_n(T^2)$, the torus braid group on $n$ strands. When we include  
the relations $s_i^2=1$,
the more complicated relations in $B_n(T^2)$ collapse (see  
\cite{Birman69} for
a presentation of $B_n(T^2)$)
leaving only the relations we have supplied here.

Firstly, we have $SL(2,\ZZ)$ generated as usual by $\{S,T\}$, note that  
$S^2$ is non-trivial
\be
S^{4}=1,\ S^{2}T=T S^{2},\ (ST)^{3}=1.
\ende
The braid group has only $n-1$ permutations but after imposing  
$s_i^{2}=1$ we can build one
more permutation which enhances the symmetric group into the  
$\widehat{A}_{n-1}$ affine Weyl group,
\bea
&& s_n=t_1^{-1} t_n s_1 s_2 \ldots s_{n-1}, \\
&& s_i^2=1,\\
&& s_{i}s_{i+1}s_{i}=s_{i+1}s_{i}s_{i+1} \ \forall i\  (n\neq  
2)\label{affone}\\
&& s_{i}s_{j}=s_{j}s_{i},\ |i-j|>1. \label{afftwo}
\eea
Taken together \ref{affone} and \ref{afftwo} give the canonical  
presentation of $\widehat{A}_{n-1}$.
Other relations within the braid generators are
\bea
&& t_iT_j=T_{j}t_{i}, \\
&& s_i t_i s_i=t_{i+1},\ s_i t_{i+1}s_i=t_{i}, \\
&& s_i T_i s_i=T_{i+1},\ s_i T_{i+1}s_i=T_{i} \ \forall i.
\eea
As mentioned above, the mapping class group is an extension of  
$B_n(T^2)$ by $SL(2,\ZZ)$,
explicitly this is given by
\bea
&& S^{-1}s_iS=s_{i}\ i\neq n, \\
&& T^{-1}s_iT=s_{i},\ i\neq n,\\
&& S^{-1}T_i S= t_i,\ S^{-1}t_i S= T_i^{-1},\ \forall \ i,\\
&& T^{-1}t_iT= T_it_i,\ T^{-1}T_iT=T_i \ \forall \ i.
\eea
\end{appendix}

\end{document}